\documentclass[12pt,preprint]{aastex}

\slugcomment{To appear in Ap. J.}

\shorttitle{Simulated Characteristics of Chromospheric Evaporation}
\shortauthors{Brannon \& Longcope}

\begin{document}


\title{Modeling properties of chromospheric evaporation driven by thermal conduction fronts from reconnection shocks}


\author{Sean Brannon and Dana Longcope}
\affil{Department of Physics, Montana State University, Bozeman, MT 59717, USA}




\begin{abstract}
Magnetic reconnection in the corona results in contracting flare loops, releasing energy into plasma heating and shocks.  The hydrodynamic shocks so produced drive thermal conduction fronts (TCFs) which transport energy into the chromosphere and drive upflows (evaporation) and downflows (condensation) in the cooler, denser footpoint plasma.  Observations have revealed that certain properties of the transition point between evaporation and condensation (the ``flow reversal point'' or FRP), such as temperature and velocity-temperature derivative at the FRP, vary between different flares.  These properties may provide a diagnostic tool to determine parameters of the coronal energy release mechanism and the loop atmosphere.  In this study, we develop a 1-D hydrodynamical flare loop model with a simplified three-region atmosphere (chromosphere/transition region/corona), with TCFs initiated by shocks introduced in the corona.  We investigate the effect of two different flare loop parameters (post-shock temperature and transition region temperature ratio) on the FRP properties.  We find that both of the evaporation characteristics have scaling-law relationships to the varied flare parameters, and we report the scaling exponents for our model.  This provides a means of using spectroscopic observations of the chromosphere as quantitative diagnostics of flare energy release in the corona.
\end{abstract}



\section{Introduction}
\label{sec:intro}

The generally accepted picture of the solar flare process begins with the reconnection of magnetic field lines in the solar corona.  The freshly reconnected flare loop is then free to retract under magnetic tension, which heats and compresses the loop-top plasma, forming hydrodynamic shocks \citep{longcope09} and accelerating electrons near the loop apex.  In the case of shocks, the steep temperature gradient between the ambient coronal plasma and the hotter post-shock plasma results in thermal conduction fronts (TCFs) that rapidly propagate down each leg of the loop \citep{craig76,forbes89,tsuneta96}.  In the case of accelerated electrons, the result is a large flux of non-thermal particles (NTPs) that precipitate down the loop towards the footpoints \citep{brown73}.

Although the question of which of these two models constitutes the dominant energy transport mechanism has not been resolved, the end result is similar; namely, the transport of energy down the loop which is subsequently deposited in the cooler and denser plasma in the transition region (TR) and chromosphere that lie at the loop footpoints.  This deposition of energy creates a significant overpressure in the TR and upper chromosphere, and drives flows of heated plasma both up and down the loop \citep{fisher87}.  These upflows and downflows are historically referred to as chromospheric evaporation \citep{sturrock73} and condensation \citep{fisher89}, respectively, and should be distinguished from unrelated flows in the corona (such as {\em coronal rain}) that are driven by thermal instabilities due to radiative losses. Finally, the evaporation of heated dense plasma from the chromosphere fills the loop and forms the bright coronal flare loops that are visible at temperatures of several million kelvin (MK).

A critical component to understanding the subsequent flare loop development is a detailed knowledge of the characteristics of chromospheric flows during a flare.  One tool for determining these characteristics are observations of Doppler spectral line shifts, which give the flow velocities within the loop at different plasma temperatures.  Ideally, the Doppler line shifts would be observed with sufficient resolution (in temperature) to give a velocity profile near the point separating upflows from downflows, which would be useful in constraining the mechanism driving the flows.  Unfortunately, it has proven difficult to obtain this data.  Most studies instead prefer to investigate evaporation using only a small set of spectral lines, sometimes only a single one \citep{czaykovska99,wulser94,zarro89}.  Other studies concentrate instead on using observations to calculate other properties of the flare loop, such as the total quantity of evaporated plasma \citep{acton82}.

In the last few years, however, high-resolution spectral observations of flare footpoints have become possible thanks to the Extreme-ultraviolet Imaging Spectrometer (EIS) located onboard the \emph{Hinode} spacecraft.  The large number of spectral lines available with this instrument has allowed for Doppler-shifts to be derived for plasma across a broad temperature range for several different flares \citep{milligan06, milligan09, milligan11, li11}.  Inspection of the velocity-temperature data in these papers reveals three interesting features.  First, the temperature of the point separating upflows from downflows (which we dub the ``flow reversal point'' or FRP) is at or above 1 MK but varies widely among observed flares (in \cite{li11} it was well over 6 MK), hinting at a possible connection to properties of the flare loop.  Second, the flows are broadly distributed in temperature, from $10^5$ K for downflows to over $10^7$ K for upflows.  Finally, these studies used a sufficient number of spectral lines to allow a rough calculation of the velocity derivative with respect to plasma temperature, which like the flow conversion temperature varies between different flares.

Significant effort has also been devoted to modeling chromospheric evaporation using computer simulations.  These simulations have generally invoked one of two candidates for energy transport: non-thermal particle (NTP) precipitation \citep{macneice84,nagai84,fisher85}, and thermal conduction front (TCF) heating \citep{nagai80,cheng84,macneice86,fisher86}.  NTP models have had some success explaining flow observations.  However, no model yet exists which is capable of self-consistently tracking the conversion of magnetic energy, released by reconnection, into a population of NTPs.  Lacking this feature, simulations must resort to introducing the non-thermal electrons {\em ad hoc}, with a user-specified energy flux and spectrum.  Properties of the evaporation flows, such as flow conversion temperature, naturally depend on this {\em ad hoc} choice.

The case of TCFs is notably different owing to the existence of comprehensive models of reconnection energy release.  Large scale models of reconnection, such as the early model of \cite{petschek64}, have used hydrodynamic equations and thus omitted energetically significant non-thermal populations.  In these models, kinetic energy is converted to thermal energy at MHD shocks, raising the post-shock loop top plasma temperature and originating TCFs due to steep temperature gradients.  It is therefore possible to use these models to study how the properties of chromospheric evaporation depend on the magnetic reconnection providing the energy.  As yet, however, there are no generally accepted relationships that predict evaporation velocities from flare energy.  Now that observations of footpoint velocities during a flare have been made with sufficient detail to determine characteristic properties of the flows, we wish to use this characterization to infer the properties of coronal energy release.

In this paper, we use a numerical simulation code to investigate the relationship between observable properties of chromospheric evaporation during a flare and the initial properties of the flare loop.  Our goal is to systematically cover a parameter space of simulation inputs in order to extract a scaling-law relationship with the output observed quantities.  First, in Section \ref{sec:model}, we develop a simplified model of a flare loop, reducing the more complex 2-D dynamics to a 1-D shocktube.  In Section \ref{sec:code} we describe the details of our numerical simulation code, including our simplified loop atmosphere model and shock initialization.  Then, in Section \ref{sec:sim}, we detail the evolution of one particular simulation including the basic hydrodynamics and the differential emission measure, develop a consistent method of extracting synthetic Doppler velocities similar to observations, and compare the results to one particular set of observed flow velocities.  Finally, in Section \ref{sec:powerlaws} we describe the parameter survey we use to extract scaling law relationships between inputs and synthetic observations, and determine the best-fit parameters.

\section{Flare Loop Model}
\label{sec:model}

\begin{figure}
\epsscale{0.5}
\plotone{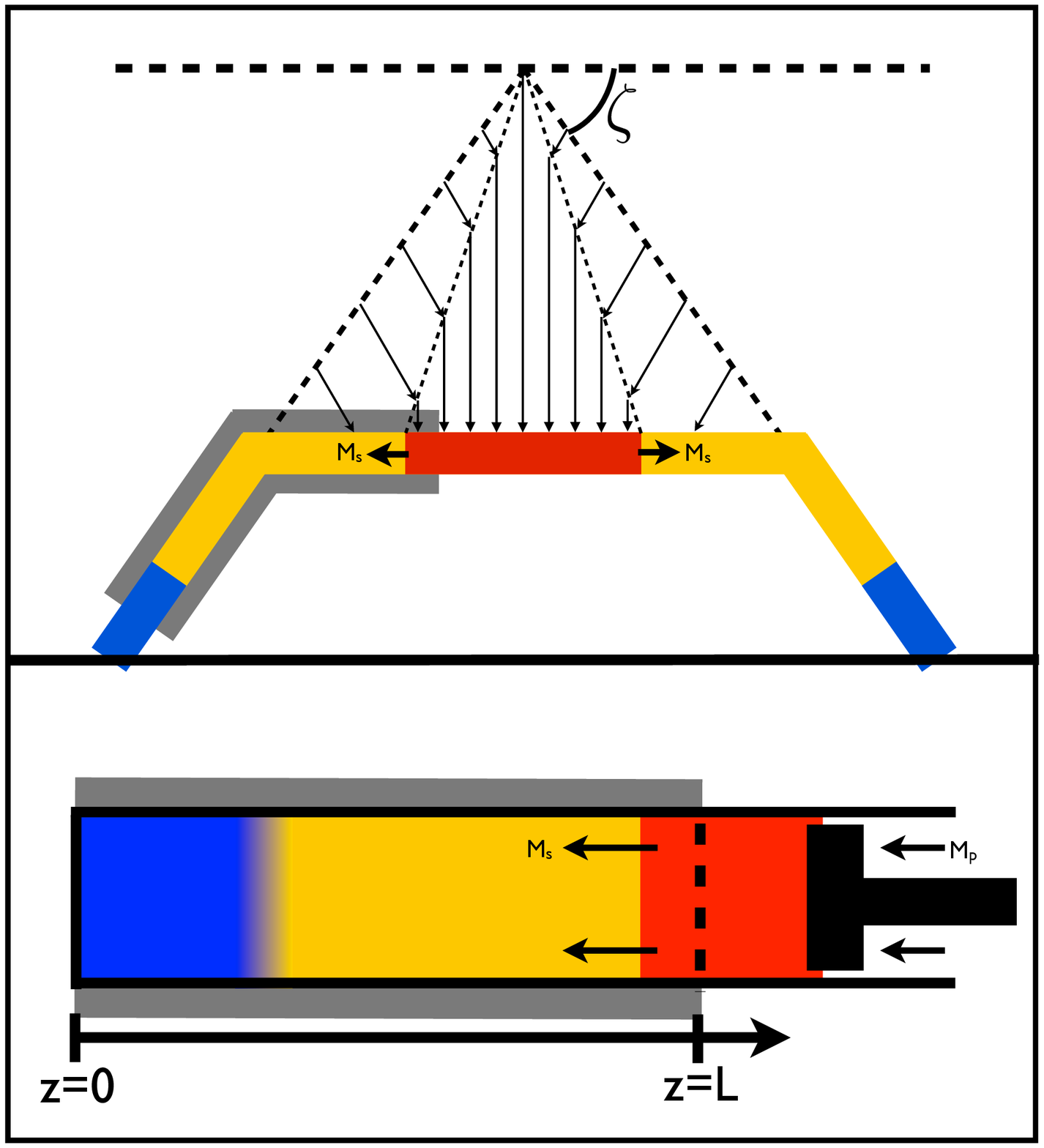}
\caption{Top: schematic diagram of a reconnected flare loop, with initial flux tube geometry shown by the long dashed lines and the reconnection angle $\zeta$ is defined as indicated.  Colored portion shows the later flux tube position after contraction, with yellow and blue respectively indicating coronal and chromospheric plasma.  Solid arrows indicate trajectory of accelerated coronal plasma before and after the slow-mode shocks (short dashed line), and the red region indicates the hot, compressed post-shock plasma.  Bottom: schematic diagram of the simplified ``shocktube'' model used in this paper, after neglecting gravity and loop geometry (color-coding is identical).  The gas dynamic shock is driven by an assumed piston (far right) moving leftward at $M_p$.  The gray box in both schematics indicates the simulation region.
\label{fig:loop_model}}
\end{figure}

The idealized flare loop model we use in this paper is an extension of the thin-flux-tube model developed in \cite{longcope09}.  In that model a brief, localized reconnection event is assumed to have occurred between two adjacent magnetic flux tubes previously separated by a current sheet.  The sheet exists between field lines whose directions differ by less that 180$^{\circ}$ (i.e.\ field which is not perfectly anti-parallel).  The result is a ``$\Lambda$''-shaped loop such as the one shown in the upper schematic in Figure \ref{fig:loop_model} (adapted from Figure 2 in \cite{longcope09}) by the long dashed lines.  In this picture the current sheet is located above the dashed line in the plane of the diagram, and the reconnection angle $\zeta$ between the field lines is defined as indicated.  This angle, apparent when viewing the current sheet from the side, differs from the narrow opening angle between shocks when the sheet is viewed edge-on.  The latter angle has been a focus of steady-state modeling such as the seminal work of \cite{petschek64}; this is {\em not} the angle $\zeta$.  Also note that a similar ``V''-shaped field line would also have resulted from the reconnection, however we omit that portion in our schematic.  This model also assumes that the ratio of gas pressure to magnetic pressure, known as the plasma-$\beta$ parameter and defined by 
\begin{equation}
\beta = \frac{8 \pi p}{B^2}
\end{equation}
where $B$ is the magnetic field strength, is much less than unity.  This is generally true of the pre-flare corona and transition region (TR) \citep{gary01}.  The plasma-$\beta$ will have increased in the retracting flux tubes (i.e.\ outflow jets), but provided the reconnecting field was sufficiently far from anti-parallel it will still be less than unity \citep{longcope09}.   Within the TCFs, which will be our primary concern, $\beta$ will lie between the initial value and that of the compressed, heated loop-top.  Under the assumption of small $\beta$, both the plasma and the thermal conductive flux are constrained to move only along the field line.  It is also assumed in order to justify a one-dimensional treatment that the tube of reconnected flux is ``thin'' in the sense that the scale of variations along the loops are generally much greater than their widths.  Finally, in addition to the background model developed in \cite{longcope09}, we introduce a cool, dense chromosphere at the feet of the loop (shown as the blue portion of the tube in the upper schematic in Figure \ref{fig:loop_model}), which acts as a mass reservoir.

After the initial reconnection event the subsequent contraction of the field line under magnetic tension results in a shorter loop shown by the colored portion in the schematic, with the ambient coronal plasma indicated in yellow.  As the loop contracts, free magnetic energy is released into accelerating the plasma downward and inward; the inward motion corresponds to motion parallel to the axis of the flux tube \citep{longcope09}.  Starting from the initial configuration, as the contracting loop passes each of the angled solid arrows the plasma at that location is accelerated by the rotational discontinuity down and inward toward the loop center; the subsequent trajectory of the plasma is the solid arrow itself.  Eventually this accelerated plasma piles up at the loop top, as shown by the red region in the schematic, resulting in heating and compression of the plasma and the formation of two slow magnetosonic shocks resembling simple gas dynamic shocks.  These shocks propagate out along the loop at a hydrodynamic Mach number $M_s$, determined in terms of the reconnection angle $\zeta$ by \cite{longcope09} as
\begin{equation}
M_{s} = \sqrt{\frac{8}{\gamma \beta}} \sin^{2} \left( \zeta / 2 \right),
\end{equation}
and they follow the trajectories given by the short dashed lines.  Note that the effect of the shocks is to alter the flow of the loop plasma from downward and inward to purely downward motion \citep{longcope09}.  At the same time, strong temperature gradients across the shock fronts, going from multi-MK post-shock plasma to $\sim$1 MK in the pre-shock coronal plasma, give rise to fast-moving thermal conduction fronts (TCFs) which move out along the loop ahead of the shocks.

Since our interest is in the effects of a shock-initiated TCF on the chromosphere, and not in the overall dynamics of the loop evolution, we narrow our focus to only that section of the flare loop indicated by the gray box in the upper schematic of Figure \ref{fig:loop_model}.  We also adopt a reference frame that is co-moving with the contracting loop, so that the ambient coronal plasma (yellow) is stationary and the post-shock plasma (red) is being driven down the loop.  We further simplify the model by neglecting gravitational stratification of the plasma.  The resulting horizontal ``shocktube'' model of the flare loop is shown in the lower schematic in Figure \ref{fig:loop_model}, with the region of interest again indicated in gray.  In this model, the post-shock plasma behaves as though driven by a piston, located to the right of the region of interest and moving leftward at a Mach number $M_p$ (referring to the pre-shock coronal plasma), and the shock front moves leftward down the tube at $M_s$.  Finally, we include a simplified model of the TR and chromosphere (blue in the schematic), the details of which will be discussed in Section \ref{sec:atmosphere}.

\section{Simulation Setup}
\label{sec:code}

\subsection{1-D Fluid Equations}
\label{sec:equations}

Following the above discussion we consider a one-dimensional shocktube of plasma with uniform cross-section and total length $L$, parameterized by a coordinate $0 \le z \le L$, as shown in the lower schematic of Figure \ref{fig:loop_model}.  We wish to numerically simulate the plasma hydrodynamics within the tube, beginning at an initial time $t_0=0$ forward to some later time $t$.  We begin by assuming that the plasma is everywhere of sufficient collisionality to be adequately described as a single-fluid with pressure $p$, proton number density $n$, average flow velocity $v$, and temperature $T$.  In this case, we recall the 1-D hydrodynamic equations for an ideal fluid, given by
\begin{equation}
\frac{\partial n}{\partial t} = - \frac{\partial}{\partial z} \left[ n v \right];
\label{eq:mass}
\end{equation}
\begin{equation}
\frac{\partial v}{\partial t} = - v \frac{\partial v}{\partial z} - \frac{1}{m_p n} \left( \frac{\partial p}{\partial z} - \mu \frac{\partial^2 v}{\partial z^2} \right);
\label{eq:momentum}
\end{equation}
\begin{equation}
\frac{\partial T}{\partial t} = - v \frac{\partial T}{\partial z} - \left( \gamma - 1 \right) T \frac{\partial v}{\partial z}  + \frac{\gamma - 1}{k_b n}  \left\lbrace \frac{\partial}{\partial z} \kappa \frac{\partial T}{\partial z} + \mu \left| \frac{\partial v}{\partial z} \right|^{2} + \dot{Q}^{(ext)} \right\rbrace,
\label{eq:energy}
\end{equation}
where $m_p$ is the proton mass, $k_b$ Boltzmann's constant, $\mu$ is the parallel dynamic viscosity, and $\kappa$ the thermal conductivity (discussed in Section \ref{sec:visccond}).  We adopt gas constant $\gamma = 5/3$ for a fully ionized monatomic plasma.  Note that we do not include gravity in Equation (\ref{eq:momentum}), and hence we neglect gravitation stratification.  We also do not treat explicit coronal heating or plasma radiation in Equation (\ref{eq:energy}), and instead have included a single heating/cooling source term $\dot{Q}^{(ext)}$ (discussed in Section \ref{sec:atmosphere}) that is responsible for the equilibrium loop atmosphere.  Finally, we close the system with the ideal gas law,
\begin{equation}
p = 2 k_b n T.
\label{eq:ideal}
\end{equation}

For this study, we define a system of dimensionless variables, where the coronal number density $n_{cor}$, temperature $T_{cor}$, sound speed $c_{s,cor}$, and proton mass $m_p$ are scaled to unity.  From the equation for sound speed, $c_{s} = \sqrt{\gamma p / m_p n}$, we see that the coronal pressure is rescaled to $p_{cor} = 0.6$.  Length $z$ is rescaled by the coronal ion mean free path, given by
\begin{equation}
\ell_{mfp}^{(cor)} = \frac{4}{3} \frac{\mu}{m_p n c_{s}} = 58.5 \mbox{ km} \left( \frac{T_{cor}}{1 \times 10^{6}} \right) \left( \frac{1 \times 10^{9}}{n_{cor}} \right)
\label{eq:mfp}
\end{equation}
after using the classical Spitzer viscosity \citep{spitzer53}, and time $t$ is rescaled to the sound transit time $\ell_{mfp} / c_{s,cor}$.  Note that these new variables do not alter the form of Equations (\ref{eq:mass})-(\ref{eq:ideal}), except that $k_{b}$ is formally replaced by $1/2\gamma$ via Equation (\ref{eq:ideal}).  Throughout the remainder of this section, we shall assume the use of the dimensionless variables.

\subsection{Numerical Integration}
\label{sec:integration}

To numerically integrate the hydrodynamic Equations (\ref{eq:mass})-(\ref{eq:energy}), we first construct a staggered grid $G_i$ of total length $L=100 \cdot \ell_{mfp}$ and uniform cell size $\Delta z = 0.05$ which defines the simulation region.  The total size of the grid defined in this way is 2000 cells, to which we add two additional sets of static cells on either end to enforce the boundary conditions.  These static cells are reset to their initial values after each time step.  The lower boundary $z=0$ is completely closed ($v=0$, $\kappa=0$), and the treatment of the upper boundary will be discussed in Section \ref{sec:shock}.  The values for the hydrodynamic variables are defined at each point on the staggered grid: bulk quantities such as $p$ and $\mu$ are defined at cell centers, and flux quantities such as $v$ and $\kappa$ are defined at cell edges.  We have tested our code using both 2000 and 4000 cells and found that the results do not substantially differ.  We have also tested that the staggered scheme conserves mass, momentum, and energy over the simulation region, which it generally does to within $\pm$0.1\% during the simulation.

With the grid and fluid variables defined, we numerically integrate Equations (\ref{eq:mass})-(\ref{eq:energy}) using an explicit midpoint-stabilized stepping-algorithm for all terms except for thermal conductivity in Equation (\ref{eq:energy}).  Were a fully explicit scheme used the timestep size $\Delta t$ would be chosen to satisfy the Courant conditions \citep{courant67},
\begin{equation}
\Delta t \le \min \left( \frac{\Delta z_i}{c_{s,i}}, \, \frac{\Delta z_i}{v_i}, \, \frac{ n_i \Delta z_{i}^2}{\mu_i}, \, \frac{n_i \Delta z_{i}^2}{\gamma \left( \gamma -1 \right) \kappa_i} \right)
\label{eq:courant}
\end{equation}
where the minimum is taken over the full set of grid points $G_i$.  The first two conditions are the sound wave and flow velocity timescales, and the third is the viscous timescale.  The final condition is the conductive timescale, which is in general significantly smaller than any of the other three.  This is because the Prandtl number, which defines the ratio of viscosity to the thermal conductivity, is typically of order $Pr \sim 0.01$ for a plasma.  This results in a conductive timescale that is at least 100 times smaller than any of the other timescales, and also results in prohibitive runtimes for an explicit numerical code.

We circumvent this issue by first expanding the thermal conductive term in Equation \ref{eq:energy} as
\begin{equation}
\left( \frac{\partial T}{\partial t} \right)_{cond} = \frac{\gamma \left( \gamma - 1 \right)}{n} \frac{\partial}{\partial z} \kappa \frac{\partial T}{\partial z} = \frac{\gamma \left( \gamma - 1 \right)}{n} \left\lbrace \frac{\partial T}{\partial z} \frac{\partial \kappa}{\partial z} + \kappa \frac{\partial^2 T}{\partial z^2} \right\rbrace,
\end{equation}
and then implementing an implicit \emph{Crank-Nicolson} integration method \citep{crank47} for the second-derivative term (the first term is folded into the normal explicit solver).  This semi-implicit scheme permits us to effectively ignore the conductive Courant condition and use only the minimum of the first three terms in Equation (\ref{eq:courant}).  As the numerical integration proceeds, the values for $p$, $v$, $n$, and $T$ for the entire grid are saved every $t_{frame} = 0.01$.  The entire simulation is allowed to run until the thermal conduction front, which begins at the top of the tube and propagates down, reaches the lower boundary, at which point the closed boundary condition would begin reflecting waves back up the tube.  As this represents undesired (and possibly unphysical) behavior, the simulation is ended at that time.

\subsection{Viscosity and Conductivity}
\label{sec:visccond}

A major obstacle to keeping the hydrodynamics well-resolved in any flare loop simulation that includes both the corona and the chromosphere is the fact that the ion mean free path given in Equation \ref{eq:mfp}, which governs the length scale over which hydrodynamic quantities may vary significantly, becomes decidedly smaller as we move down from the corona into the chromosphere.  In general, the chromospheric temperature is of order 100 times lower than in the corona and the density 100 times higher.  Using the standard Spitzer formula for viscosity $\mu=\mu_0 T^{5/2}$ \citep{spitzer53}, and noting that $c_s \propto T^{1/2}$ for a plasma, then we see from Equation (\ref{eq:mfp}) that
\begin{equation}
\ell_{mfp} \propto \frac{T^2}{n},
\end{equation}
which results in a mean free path that is six orders of magnitude smaller in the chromosphere than in the corona.  For our grid spacing of $\Delta z = 0.05$ this implies that there would be $\sim$50,000 mean free paths per grid cell in the chromosphere, which is inadequate to resolve fine structure hydrodynamics such as shocks.

One popular method to circumvent this issue is to use a non-uniform adaptive grid that can add or subtract grid points of varying size during the simulation to increase resolution where needed.  Several established methods exist for running hydrodynamic simulations with adaptive grids, e.g. PLUTO \citep{mignone07}, although it is not entirely clear that such methods are able to adequately resolve shock structures in the chromosphere.  Moreover, given the 1-D low plasma-$\beta$ nature of our hydrodynamic model, there are no additional benefits to using an adaptive grid scheme.  We therefore adopt a different approach, modifying the standard Spitzer formula for viscosity by adding an additional term of the form
\begin{equation}
\mu = \mu_0 \left( T^{5/2} + \alpha n c_s \right).
\label{eq:mu}
\end{equation}
We see from Equation (\ref{eq:mfp}) that the effective mean free path then becomes
\begin{equation}
\ell_{mfp}^{(eff)} = \frac{4}{3} \mu_0 \left( \frac{T^{5/2}}{n c_s} + \alpha \right),
\label{eq:mfpmod}
\end{equation}
and in the corona, where we demand $T_{cor}$, $n_{cor}$, and $c_{s,cor}$ and now $\ell^{cor,eff}_{mfp}$ are all scaled to unity, we can solve for $\mu_0$ as
\begin{equation}
\mu_0 = \frac{3}{4 \left( 1+\alpha \right)}.
\label{eq:mu0}
\end{equation}
To determine $\alpha$, we consider the mean free path in the chromosphere.  In this region, due to the low temperature and high density, the first term in Equation (\ref{eq:mfpmod}) essentially vanishes, leaving
\begin{equation}
\ell_{mfp}^{(chr)} = \frac{4}{3} \mu_0 \alpha.
\label{eq:mfpchr}
\end{equation}
If we now impose the condition that $\ell_{mfp} \ge \ell_0$ for all points in the tube, where $\ell_0$ is a lower bound that artificially boosts the mean free path in the chromosphere, we can then solve for $\alpha$ using Equations (\ref{eq:mu0}) and (\ref{eq:mfpchr}) to obtain
\begin{equation}
\alpha = \frac{\ell_0}{1-\ell_0}.
\end{equation}
We find through experimentation that $\ell_0=0.01$ seems to result in adequate resolution of the hydrodynamics in the chromosphere, which thus sets $\alpha=1/90$ and $\mu_0 = 0.7425$.  We also performed test runs with $\ell_0=0.005$, with the observed result that shocks became poorly resolved in the chromosphere (manifested as a slowly growing sawtooth behind the shock) and a roughly 10-20\% increase in the flow reversal properties described in Section \ref{sec:wtdveltemp}.

To determine the thermal conductivity in this model, we first recall the definition of the Prandtl number,
\begin{equation}
Pr = \frac{4}{3} \frac{\mu}{k_b \kappa} = \frac{8 \gamma}{3} \frac{\mu}{\kappa};
\end{equation}
we adopt $Pr = 0.012$ for the duration of this paper.  Note that the modified version of $\mu$ in Equation (\ref{eq:mu}) would result in different Prandtl numbers for the corona and the chromosphere if we used the standard Spitzer formula for conductivity $\kappa=\kappa_0 T^{5/2}$.  Consequently, we modify the thermal conductivity in the same manner as the viscosity, with
\begin{equation}
\kappa = \kappa_0 \left( T^{5/2} + \alpha n c_s \right).
\label{eq:kappa}
\end{equation}
Substituting this expression into the expression for the Prandtl number and solving for $\kappa_0$ results in $\kappa_0=275$.

\subsection{Initial Loop Atmosphere}
\label{sec:atmosphere}

\begin{figure}
\epsscale{0.5}
\plotone{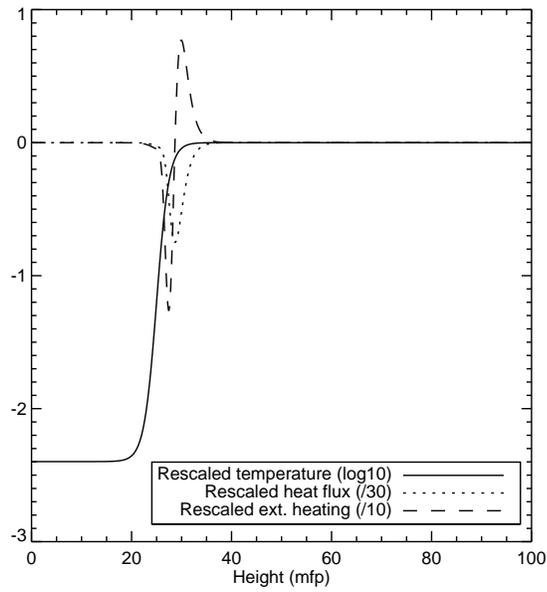}
\caption{General properties of the artificial initial loop atmosphere (chromosphere, TR, and corona) as described in Section \ref{sec:atmosphere}, in rescaled units and for a temperature ratio of $R=250$.  Solid line: log-plot of the rescaled temperature profile.  Dotted line: rescaled heat flux within the TR (divided by 30).  Dashed line: rescaled external heating (divided by 10) supplied to the atmosphere as a proxy for coronal heating (positive) and chromospheric radiation (negative).
\label{fig:atmosphere}}
\end{figure}

The left-hand portion of the simulation region contains the TR and chromosphere, as shown in Figure \ref{fig:loop_model}.  It is well known that the structure of a static TR and chromosphere depends critically on radiation, gravity, and even ionization states (e.g.\ \cite{vernazza81}, \cite{fontenla90}, etc.).  This layer responds so rapidly to the heat flux from flare reconnection, however, that these mechanisms play little role in the evaporation dynamics.  The main factor determining the dynamic response is, instead, the pre-flare distribution of mass density.  We therefore use a simplified physical model tuned to produce a relatively realistic initial density distribution.  The chief aspect we seek to reproduce is the very large ratio of temperatures and densities, which we quantify as
\begin{equation}
R = \frac{T_{cor}}{T_{chr}} = \frac{1}{T_{chr}}.
\label{eq:temp_ratio}
\end{equation}
Since we omit gravity the initial pressure is uniform and the density ratio is the inverse of the temperature ratio.  The initial distribution is given by the expression
\begin{equation}
\log_{10} T_{atmo} = \frac{1}{2} \log_{10}\left( \frac{1}{R} \right) \left[ 1 - \tanh \left( \frac{z - z_{TR}}{d} \right) \right].
\label{eq:model_atmo}
\end{equation}
where $z_{TR}$ is the center of the TR and $d$ is a measure of its thickness.  Throughout this study we shall set $z_{TR}=25$ (one-quarter of the way up the tube from the left-end) and $d=2.5$.  In Figure \ref{fig:atmosphere} we plot the temperature profile $\log_{10} T_{atmo}(z)$ for $R=250$ as the solid line, and by inspection we see that our choice of $d=2.5$ results in a TR that is $\sim$10 coronal mean free paths thick.

The steep temperature gradient across the TR naturally results in a strong thermal conductive flux, given by
\begin{equation}
F_{c} = - \kappa \frac{\partial T}{\partial z},
\label{eq:heat_flux}
\end{equation}
which transfers thermal energy from the hot corona to the much cooler chromosphere.  This thermal flux (divided by 30) for the $R=250$ atmosphere is plotted as the dotted line in Figure \ref{fig:atmosphere}, and clearly shows that the majority of the thermal flux occurs in the upper portion of the TR.  This feature is a result of the temperature profile being defined on a log-scale, which implies that the strongest gradients in the TR will be at the higher temperature.

In order to maintain the initial temperature profile, Equation (\ref{eq:model_atmo}), against the action of the thermal conductive flux we introduce an \emph{ad hoc} heating term to Equation (\ref{eq:energy}),
\begin{equation}
\dot{Q}^{(ext)} = - \frac{\partial}{\partial z} \left[ \kappa \frac{\partial T_{atmo}}{\partial z} \right].
\label{eq:heating_balance}
\end{equation}
This term (divided by 10) is plotted as the dashed line in Figure \ref{fig:atmosphere}, and consists of a source ($\dot{Q}>0$) in the upper layer and sink ($\dot{Q}<0$) in the lower layer.  These artificial elements stand in for coronal heating and chromospheric radiation, respectively.  We keep Equation (\ref{eq:heating_balance}) constant throughout the run, although we find that if we turn it off during the flare simulation there is no discernible difference in the chromospheric response.  This indicates that the heating mechanism is not an essential property in determining evaporation evolution, but rather the initial mass-temperature distribution of plasma in the TR.

\subsection{Initial Piston Shock}
\label{sec:shock}

The right-hand portion of the simulation region contains the downward-propagating piston shock, which will serve to initiate and drive the thermal conduction front (TCF) into the TR and chromosphere.  The classical picture of a piston shock, as shown in Figure \ref{fig:loop_model}, is of a plug of compressed fluid being driven at velocity $v=-M_p$ down into an ambient fluid at rest with $v=0$.  $M_p$ is the Mach number of the driving piston speed as measured in the rest fluid.  The shock itself is the interface between these two regions, and it propagates ahead of the piston at speed $M_s$ given by
\begin{equation}
M_s = \frac{M_p \left( \gamma + 1 \right)}{4} + \sqrt{ \left( \frac{M_p \left( \gamma + 1 \right)}{4} \right)^2 + 1 }.
\end{equation}
The plasma compression across the shock results in an increased post-shock pressure and density, as given by the \emph{Rankine-Hugoniot} conditions:
\begin{equation}
p_{ps} = \frac{2 \gamma M_{s}^{2} - \left( \gamma - 1 \right)}{\gamma \left( \gamma + 1 \right)},
\end{equation}
\begin{equation}
n_{ps} = \frac{\left( \gamma + 1 \right) M_{s}^{2}}{\left( \gamma - 1 \right) M_{s}^{2} + 2}.
\end{equation}
The post-shock temperature $T_{ps}$ is given by Equation (\ref{eq:ideal}).

In the classical piston shock, the jump in velocity from $v=0$ to $v=-M_p$ is instantaneous; the shock is a strict discontinuity in the fluid variables from pre-shock to post-shock.  For a numerical simulation, however, we need to construct a smooth transition for the fluid velocity and other variables across the shock, preferably over a length scale of a few mean free paths, to ensure adequate numerical resolution of the shock \citep{guidoni10}.  In this model we initialize the shock by superimposing the post-shock plasma conditions over the upper 20\% of the tube, scaled by a transition in the velocity centered at $z=90$ of the form
\begin{equation}
v \left( z \right) \sim \tanh \left[ \frac{ z - 90 }{\lambda} \right]
\label{eq:initial_shock}
\end{equation}
where $\lambda$ is the initial length scale of the shock.  We adopt $\lambda = 2.5$ for the remainder of this study, which results in an initial shock that is initially $\sim$10 mean free paths thick (which is identical but unrelated to the TR thickness).  The post-shock region is maintained at $(p_{ps}, n_{ps}, T_{ps})$ by a flux of plasma at speed $-M_p$ coming across the upper boundary ($z=100$).  This flux is the result of the boundary condition we enforce for the static cells on the right-hand side of the grid, which are reset at each timestep to the original post-shock conditions.  We have tested our code by simulating shocks in the absence of the TR and thermal conduction, and found that shocks do indeed propagate at the correct speed $M_{s}$ while remaining well-resolved due to the presence of viscosity in the shock region.

\section{Simulation Results}
\label{sec:sim}

To discuss the various features of the simulated loop dynamics, we focus first on a single simulation.  The qualitative results of the evolution of this particular simulation are similar for most of the runs performed in this study.   We consider a TR temperature ratio $R=250$ and piston Mach number $M_p = 2.0$, and assume an ambient coronal temperature of $T_{cor} = 2.5$ MK and number density $n_{cor} = 10^9$ cm$^{-3}$.  Restoring conventional dimensions to variables (assumed throughout this section) results in a total length for the simulation region $L = 36.6$ Mm, a coronal sound speed $c_{s,cor} = 270$ km s$^{-1}$, and a post-shock temperature $T_{ps} = 9.2$ MK.  The total duration of the simulation (from the initial state to the TCF reaching the lower boundary) is $t_{sim} = 38.0$ seconds.

\subsection{Hydrodynamics}
\label{sec:hd_evol}

The hydrodynamic evolution of the simulation region is shown in Figure \ref{fig:sim_evolution}; seven different times are plotted and color-coded according to the legend for pressure (upper left plot), velocity (upper right plot), number density (lower left plot), and temperature (lower right plot).  At the initial simulation time, $t=0.00$ sec (black line), as we move up the tube from $z=0$ Mm we note first the cool, dense chromosphere at $10^4$ K and density $2.5 \times 10^{11}$ cm$^{-3}$.  Beginning at $\sim$7 Mm, we encounter the artificial TR, which appears only in density and temperature and continues to $\sim$11 Mm.  Above the TR, we have the constant temperature and density corona, which occupies more than 50\% of the length of the tube.  Finally, centered between 31-34 Mm, we note the initial piston shock which accelerates the post-shock plasma to $-540$ km s$^{-1}$ (negative as the shock is propagating down the tube), and which heats and compresses the plasma to 9.2 MK and $3 \times 10^{9}$ cm$^{-3}$.

By $t=0.14$ sec (purple), we see the very rapid development of the TCF in the plasma temperature.  Within this time, the TCF has propagated down to $\sim$20 Mm and has closed nearly half the distance between the initial shock position and the TR.  We see from the density and velocity profiles in Figure \ref{fig:sim_evolution} that the shock itself has not moved downward very far (indeed, the density profile for the shock is scarcely different than at the initial time).  This is made clearer in the pressure profile where we note that the plasma pressure rises once between 20-31 Mm due to the presence of the TCF, and subsequently rises again across the shock between 31-35 Mm.  This decomposition of a shock, in the presence of thermal conduction, into a TCF and a so-called \emph{isothermal sub-shock} is common and has been observed in previous models \citep{guidoni10,longcope09}.

After $t=1.02$ sec of evolution (blue), the TCF has propagated downward far enough that it encounters the cooler and denser TR plasma, which results in two distinct effects.  First, the TCF slows dramatically due to the reduced thermal conductivity in the TR and chromosphere; indeed, the TCF takes 1 second to descend the $\sim$20 Mm between the initial piston shock and the TR, but takes another 37 seconds to clear the TR and chromosphere and reach the lower tube end.  Second, the TCF begins rapidly depositing thermal energy into the stationary TR plasma.  This rapid rise in plasma temperature in the upper TR, coupled with the stationary density profile, results in the development of a large overpressure clearly visible centered at $\sim$9 Mm.  This TR overpressure is responsible for the initiation of the chromospheric evaporation upflows (visible between 9-11 Mm) and associated condensation downflows (visible between 8.5-9 Mm).

As the simulation continues to $t=5.45$ sec (cyan), we see that the TCF has completely cleared the TR and has begun to directly heat the chromosphere.  Meanwhile, the evaporation and condensation has continued to develop and we see clearly that the upflows have higher speeds and a broader spatial distribution than the downflows.  This is a reflection of the momentum balance in the TR: the TCF deposits thermal energy to the TR but no net momentum, and as evident in the density profile the condensation is occurring in a denser region than the evaporation.  Also in the density profile at this time, we note the development of an evaporation front, located at $\sim$12 Mm, which is beginning to enhance the density of the upper TR and lower coronal regions, and a barely visible rarefaction region and condensation front in the lower TR.

We now jump forward to $t=15.0$ sec (green).  By this time, the evaporation region has grown to encompass nearly one-third of the tube, and has developed upflow speeds of $\sim$500 km s$^{-1}$.  Meanwhile, the condensation region is restricted to speeds less than 100 km s$^{-1}$.  In the density profile we now see the fully developed three-part structure of condensation (enhanced densities between 4.5-6.5 Mm), rarefaction (decreased densities between 6.5-9 Mm), and the evaporation front (which has strongly enhanced densities up to $\sim$16 Mm).  There has also begun to be significant interaction between the upward propagating evaporation front and the downward propagating subshock (centered at $\sim$23 Mm), with mildly enhanced densities in between.  Finally, at this point in the simulation, we begin to observe the direct effects of the artificially-enhanced thermal conductivity $\kappa$, which manifests as a ``shoulder'' in the TCF located between 5-6 Mm.

By $t=23.1$ sec (orange), the upflows have reached and passed the maximum speed during this simulation of $\sim$510 km s$^{-1}$.  For later times the maximum upflow speed in the tube is below this.  This is the result of the downward propagating subshock finally encountering the evaporation front and passing through it, which is especially evident in the pressure and density profiles at $\sim$19 Mm.  The strongly negative post-shock velocities thus begin to cancel the positive evaporation velocities, although the enhanced pressure that results from the combined compression of the shock and evaporation front means that some positive velocities will remain in the post-shock region.

At the end of this particular simulation, $t = 38.0$ sec (red), the TCF has fully passed through the chromosphere and has developed a distinct two-step profile due to the enhanced thermal conductivity.  However, the TCF ``shoulder'' remains somewhat below the lower-bound of the evaporation region, and thus is not likely influencing the development of the evaporating plasma.  Meanwhile, the piston subshock and evaporation front have fully passed each other, resulting in a region of highly enhanced density ($\sim$1.6$\times 10^{10}$ cm$^{-3}$) between 18-24 Mm.  The maximum upflow speed has been reduced to $\sim$440 km s$^{-1}$, and a uniform upflow speed of $\sim$130 km s$^{-1}$ has developed in the region between 18-24 Mm.

Finally, to conclude our discussion of the hydrodynamic evolution of the simulation we consider the ratio of the thermal conductive flux, Equation (\ref{eq:heat_flux}), to the free-streaming saturation limit, given in non-dimensional form by \cite{longcope10} as
\begin{equation}
F_{c}^{(fs)} = \frac{3}{2} \gamma^{-3/2} \sqrt{\frac{m_p}{m_e}} \left( n T^{3/2} \right),
\end{equation}
where $m_p / m_e$ is the ratio of proton to electron masses.  The ratio $F_{c}/F_{c}^{(fs)}$ is plotted in Figure \ref{fig:freestream} for the same times and with the same color scheme as in Figure \ref{fig:sim_evolution}.  We observe that at $t=0.0$ sec there are two peaks in the flux ratio: one for the TR centered at $10.5$ Mm, and another representing the shock centered at $32.5$ Mm.  We also note that the ratio is greater than unity for a narrow range of positions centered on the initial piston shock, indicating that the thermal flux across the shock is larger than the saturation value.  This might indicate a substantial problem were the thermal flux to remain supersaturated for the duration of the simulation.  However, by $t=0.14$ sec, we see that the flux ratio has been reduced to $<$0.4 everywhere in the tube, due to the development of the TCF discussed above which quickly smoothes out the initial steep temperature gradient.  This fast TCF and thermal flux development is characteristic of all simulations performed for this study, and we do not believe that this initial violation of the free-streaming saturation limit by the piston shock is of concern for the later tube evolution.

Later, as the TCF reaches the TR (1.02 sec), we note an enhancement of the flux ratio at $\sim$11 Mm as the TCF begins depositing thermal energy into the cooler TR plasma.  This peak slowly begins to subside as the TCF clears the TR (5.45 sec and on), although it is never fully eliminated, remaining as a small ``bump'' at $\sim$11 Mm.  Curiously, we also observe that the later evolution of the flux ratio ($t=15.0$, 23.1, \& 38.0 sec) somewhat mirrors that of the density and pressure.  This is especially notable for positions between 15-25 Mm, where we notice the density enhancement due to the interaction of the evaporation and subshock fronts mirrored as a suppression of the flux ratio.  This behavior is not indicative of any change in the thermal flux, as the TCF has long since flattened the temperature profile to nearly isothermal.  Rather, it is due to the increased density resulting in a larger saturation limit, lowering the flux ratio at those positions.

\subsection{Differential Emission Measure}
\label{sec:dem}

\begin{figure}
\epsscale{1.0}
\plotone{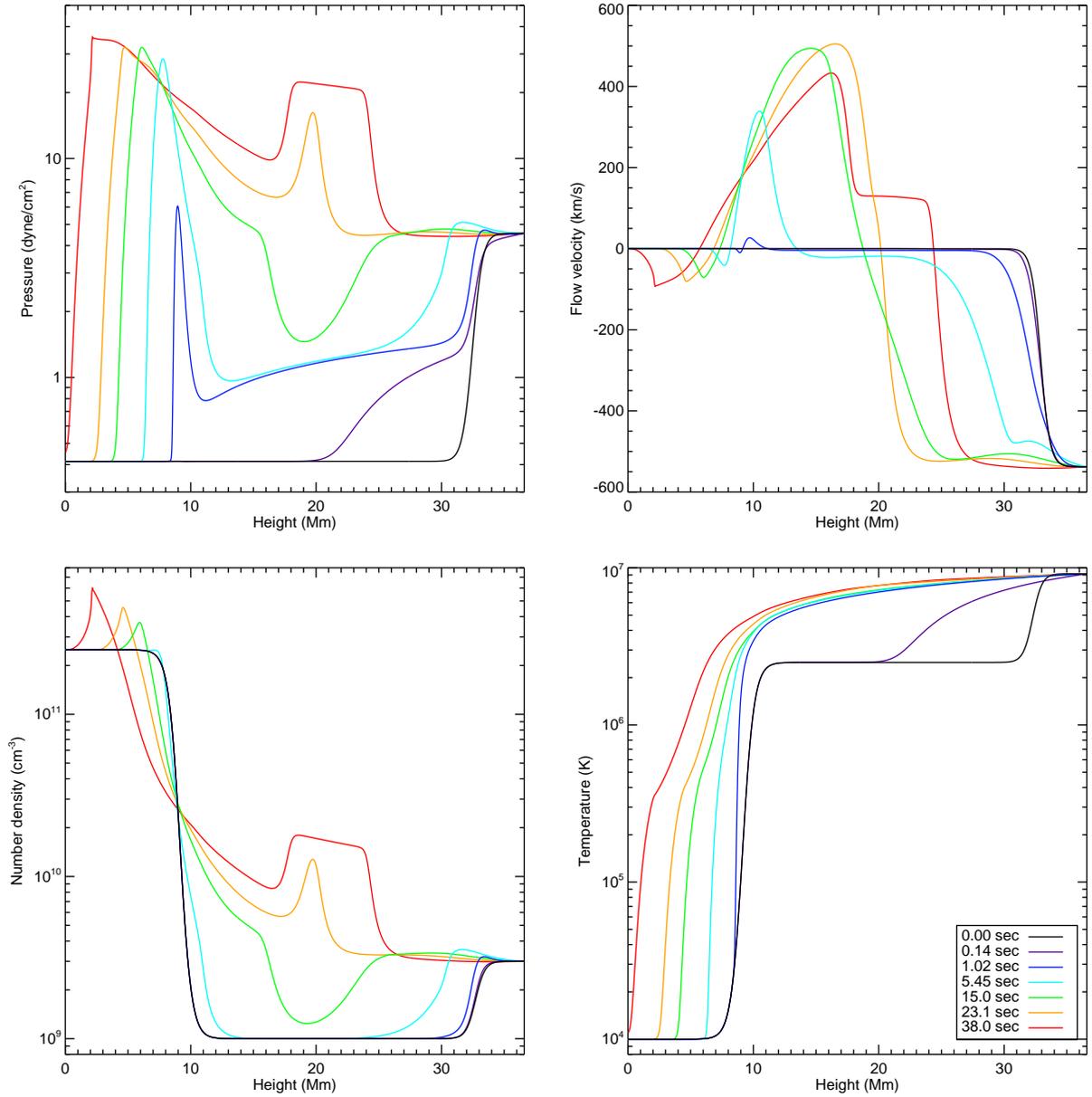}
\caption{Hydrodynamic evolution of the simulation considered in Section \ref{sec:sim} and as described in Section \ref{sec:hd_evol}.  Plasma profiles (as functions of position within the tube) are shown at seven different times during the simulation for pressure (upper left), flow velocity (upper right), number density (lower left), and temperature (lower right).  Different times are delineated by color-coding of the profiles, and the color scheme is indicated by the legend at lower right.
\label{fig:sim_evolution}}
\end{figure}

\begin{figure}
\epsscale{0.5}
\plotone{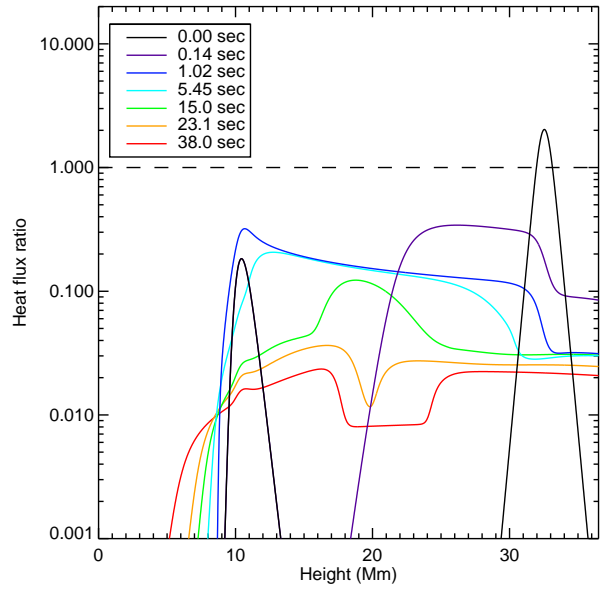}
\caption{Ratio of the conductive thermal flux to the saturated freestreaming limit during the simulation discussed in Section \ref{sec:sim}, as described in Section \ref{sec:hd_evol}.  Both the times and the color scheme are identical to Figure \ref{fig:sim_evolution}.\label{fig:freestream}}
\end{figure}

\begin{figure}
\epsscale{0.5}
\plotone{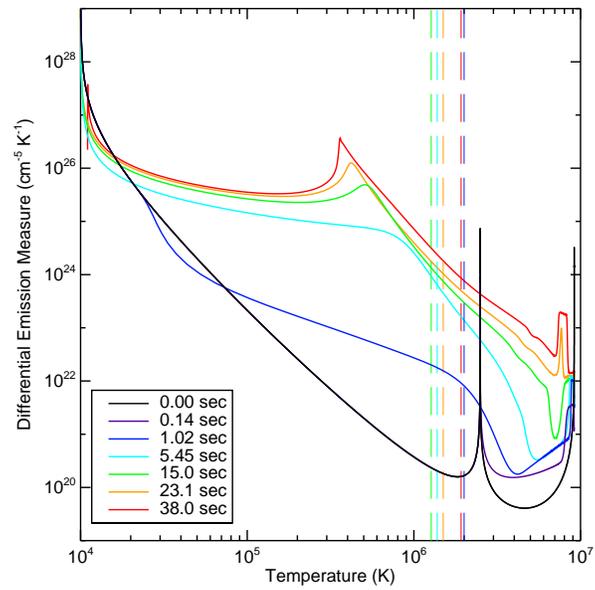}
\caption{Time evolution of the differential emission measure (DEM) for the simulation discussed in Section \ref{sec:sim} and described in Section \ref{sec:dem}, for the same times and color scheme in Figure \ref{fig:sim_evolution}.  Vertical dashed lines indicate the flow reversal point (FRP) temperature, where $v=0$ between the evaporation and condensation regions.
\label{fig:dem_evolution}}
\end{figure}

Although the full hydrodynamic evolution (as shown in Figure \ref{fig:sim_evolution} and described above, for example) would be the preferred method to understand the plasma dynamics in a flare loop, we are limited by observational techniques in our ability to extract information about those quantities.  One observational method of tracking the plasma evolution, which combines information about the density and temperature, is the \emph{differential emission measure} (DEM), defined as
\begin{equation}
DEM(T) = n_e^2 \left| \frac{dT}{dz} \right|^{-1},
\end{equation}
where $n_e$ is the electron number density (identical to the proton number density $n$ in our fully-ionized hydrogen plasma).  In Figure \ref{fig:dem_evolution} we show the evolution of the DEM as a function of temperature in the tube, for the same times and with the same color scheme as in Figures \ref{fig:sim_evolution} \& \ref{fig:freestream}.

At the initial time, $t=0.00$ sec, the clearest features of the DEM are the three sharp spikes located at $10^{4}$ K, 2.5 MK, and 9.2 MK.  These peaks correspond to the uniform temperature chromosphere, corona, and post-shock regions seen in Figure \ref{fig:sim_evolution}.  Also notable is the DEM minimum located at $\sim$ 2 MK which is a somewhat higher temperature than seen in other observational and modeled DEMs, although the overall magnitude of our DEM is comparable \citep{emslie85,brosius96}.  We attribute this higher-temperature minimum to the fact that our model atmosphere, Equation (\ref{eq:model_atmo}), has its steepest gradient $dT/dz$ at higher ($\sim$2 MK) temperatures, thus resulting in the DEM being minimized at those temperatures.

By $t=0.14$ sec, the TR portion of the DEM between $10^4$ K and 2.5 MK remains unchanged.  Only the portion of the DEM between the uniform corona and the post-shock region has been altered as the TCF begins to smooth the temperature gradient across the piston shock, resulting in some enhancement of the DEM in the 3-8 MK range.  For $t=1.02$ sec, however, we observe significant changes to the DEM in the corona and upper TR; indeed, the entire range from $10^5$ K to 2+ MK has been enhanced by a factor of 10 to 100.  At this same time, recall from Section \ref{sec:hd_evol} that upflows and downflows in the TR are beginning to form.  To differentiate between the portion of the DEM that concerns upflows from the portion that concerns downflows, we have plotted several vertical dashed lines, in the same color palette, that indicate the temperature where $v=0$ (i.e.\ the flow reversal point (FRP) separating the condensation and evaporation regions).  This will be referred to henceforth as the FRP temperature $T_{frp}$, which is defined in terms of the simulation as the temperature at the first position $z_0$ where $v \ge 0.01$.

For the $t=1.02$ sec profile, the evaporation temperatures range from the dashed blue line at 2.0 MK to slightly less than 4 MK.  Similarly, for the $t=5.45$ sec and $t=15.0$ sec profiles, the evaporation begins at $T_{frp}$ and ranges up to $\sim$5 MK and $\sim$7 MK respectively.  In these two profiles, we note the evaporation front closing with the high-temperature post-TCF subshock, represented here by the peak in the DEM at $\sim$$10^7$ K.  Note that the post-shock DEM enhancement is roughly 10-fold, and since $DEM(T) \propto n^2$ this corresponds as expected to the three-fold density increase across the subshock (see Figure \ref{fig:sim_evolution}).   We also note, in the $t=15.0$ sec profile, the DEM enhancement of the condensation front at $5 \times 10^5$ K, which is partly due to the enhanced post-condensation density and partly to the ``shoulder'' which forms on the TCF as described in Section \ref{sec:hd_evol}.

After the evaporation front and subshock interact ($t=23.1$ sec and $t=38.0$ sec profiles), we observe another roughly 10-fold DEM increase in the 7-9 MK range from the combined compression of the plasma.  Further, the continued compression of the post-condensation front plasma has continued to enhance the DEM, forming a large peak between $3 \times 10^5$ K and $4 \times 10^5$ K.  Finally, to conclude our description of the DEM evolution for this simulation, we observe that the flow conversion temperature $T_{frp}$ first appears at a somewhat higher temperature of 2.0 MK, subsequently descends to a lower range of 1.3 to 1.4 MK, and later rises again by the end of the simulation to 1.9 MK.  Although the exact temperatures vary, this decreasing-and-increasing behavior is typical of the simulations used in this study.

\subsection{Synthetic Doppler Velocities}
\label{sec:wtdveltemp}

As discussed in Section \ref{sec:intro}, observations of flare loops using the \emph{Hinode/EIS} instrument have revealed temperature-dependent Doppler velocity profiles for plasma at the loop footpoints during chromospheric evaporation.  We would thus like to construct a similar velocity-temperature profile for the simulation results, in order to compare to these observations.  However, we cannot simply use the velocity and temperature profiles as shown in Figure \ref{fig:sim_evolution}, as this is not really what is being observed by \emph{Hinode/EIS}.  Instead, the temperature-dependent Doppler velocities are derived from spectral lines from an exposure recorded over approximately 5-10 seconds, and which are weighted by the amount of emission coming from the plasma at that temperature.

We thus wish to construct a ``synthetic'' Doppler velocity to compare with data.  We begin by defining the plasma emission measure
\begin{equation}
EM = DEM(T)dT = n^2 dz,
\end{equation}
and an emission-weighted plasma velocity
\begin{equation}
v_{EM} = EM \cdot v.
\end{equation}
We next define a binned log-temperature scale with 100 equal bins per unit interval in $\log_{10} \left( T/T_{cor} \right)$, and create binned versions of the emission measure, $EM_{bin}$, and EM-weighted velocity, $v_{bin}$, by summing $EM$ and $v_{EM}$ over each temperature bin.  If there are no grid points in a given bin at that time, the values of $EM_{bin}$ and $v_{bin}$ for that bin are set to zero.  Finally, we define a time-window $W=t_{sim}/4$ and construct the synthetic Doppler velocity as
\begin{equation}
\widetilde{v}(\widetilde{T},t) = \frac{\displaystyle{\sum_{t'=t-W/2}^{t+W/2} v_{bin}(\widetilde{T},t') }}{\displaystyle{\sum_{t'=t-W/2}^{t+W/2}EM_{bin}(\widetilde{T},t')}}.
\end{equation}
The variable window size allows us to consistently accommodate different simulation durations.  For the simulation discussed thus far $W = 9.5$ sec, which is a typical exposure duration for \emph{Hinode/EIS}.  Note of course that $\widetilde{v}(\widetilde{T},t)$ is only defined for $(W/2) \le t \le (t_{sim} - W/2)$, which for the Section \ref{sec:hd_evol} simulation corresponds to times between 4.75 sec and 33.25 sec.

\begin{figure}
\epsscale{1.0}
\plotone{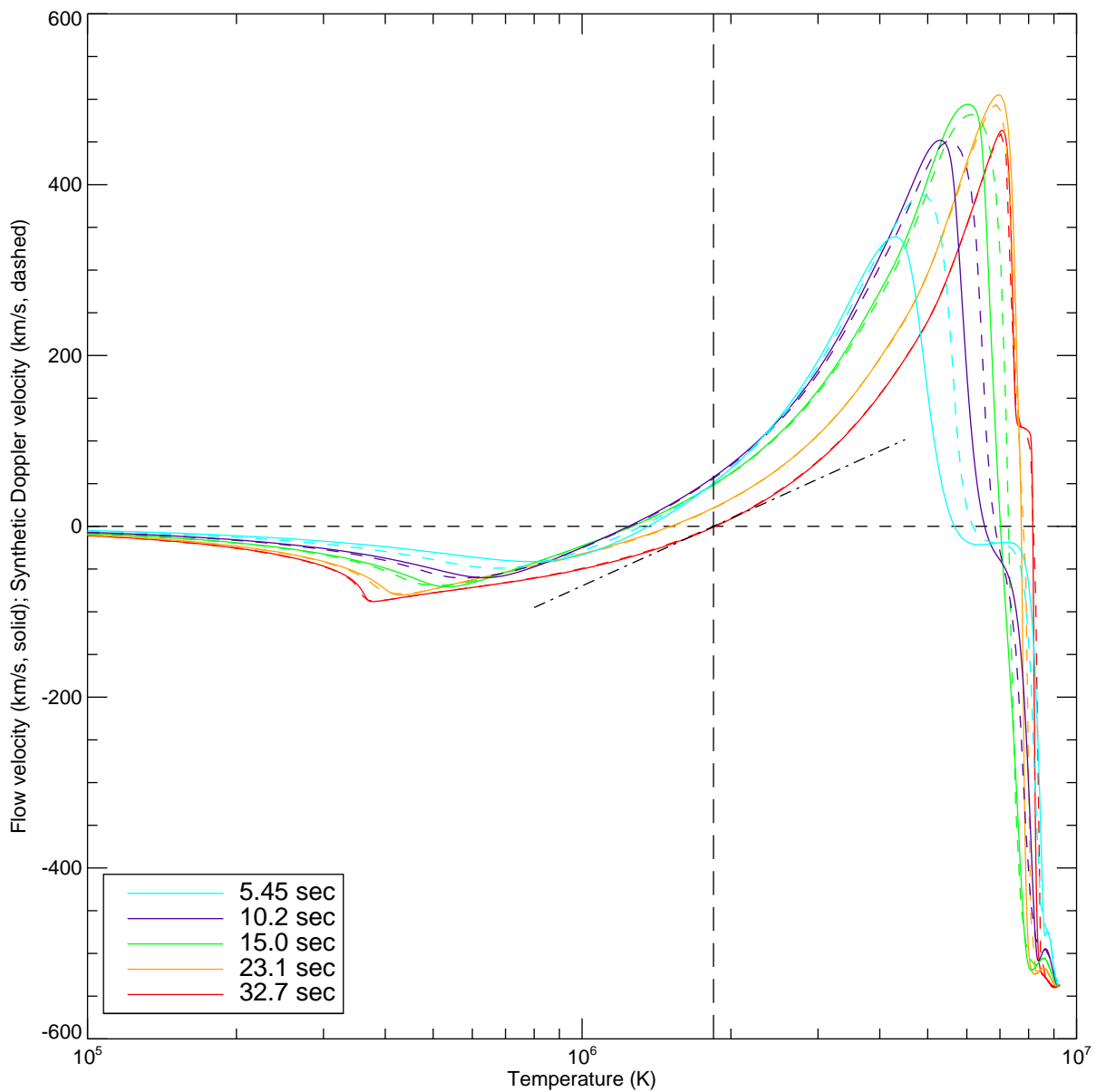}
\caption{Plasma flow velocity as a function of temperature (solid lines) and synthetic Doppler velocity as a function of binned-temperature (dashed lines) for five different times during the simulation discussed in Section \ref{sec:sim}.  Note that the times and color-coding indicated by the legend (bottom, lower left) are not identical to those in Figures \ref{fig:sim_evolution}-\ref{fig:dem_evolution}.  Horizontal dashed line indicates zero velocity.  Vertical dashed line and diagonal dash-dot line indicate the FRP temperature and slope, respectively, of the 32.7 sec synthetic Doppler profile.
\label{fig:veltemp}}
\end{figure}

In Figure \ref{fig:veltemp} we have plotted the hydrodynamic velocity $v$ as a function of temperature $T$ (solid lines) and the synthetic Doppler velocity $\widetilde{v}$ as a function of temperature $\widetilde{T}$ (dashed lines), for five representative times during the simulation and for temperatures above $10^5$ K.  Note that neither the times nor the color palette are identical to those plotted in Figures \ref{fig:sim_evolution}-\ref{fig:dem_evolution}; this is because $\widetilde{v}$ is not defined for the three earliest times or the final time in those plots.  However, we are able to include $t=5.45$ sec (cyan) which is close to the start of the windowing, $t=15.0$ sec (green), and $t=23.1$ sec (orange).  We have also added two additional times: $t=10.2$ sec (replacing purple), and $t=32.7$ sec (replacing red) which is close to the end of the windowing.

We note that one effect of the synthetic Doppler processing is to shift the velocity profiles in the upflow region to higher temperatures, particularly notable for the three earliest times.  The peak Doppler upflow speed is increased by $\sim$50 km s$^{-1}$ over the hydrodynamic velocity for the 5.45 sec profile, but is generally the same or slightly reduced for later times.  Downflow speeds are slightly increased for temperatures below 1 MK, again more significantly for early times, but the observed downflow speeds remain significantly less than the upflow speeds.  The FRP temperature $T_{frp}$ separating upflows and downflows appears mostly unaffected by the processing, however the way in which we define the FRP needs to be modified due to the temperature binning.  Recall that we previously defined $T_{frp}$ as the temperature at the first position $z_0$ where $v \ge 0.01$; we now determine the temperature bin $B_{0}$ where $\widetilde{v} \ge 0.01$ for each time $t$, and perform a fit to the two bins $\left( B_{0}, B_{0}-1 \right)$ of the form
\begin{equation}
\widetilde{v} = \widetilde{C}_0 + \widetilde{S}_{frp} \log_{10} \widetilde{T}.
\end{equation}
We now define the FRP temperature as 
\begin{equation}
\widetilde{T}_{frp} = 10^{-\widehat{C}_{0}/\widehat{S}_{frp}},
\label{eq:tfrp}
\end{equation}
and we note that $\widetilde{S}_{frp}$ is the slope of the velocity-temperature profile at the FRP.

\begin{figure}
\epsscale{0.5}
\plotone{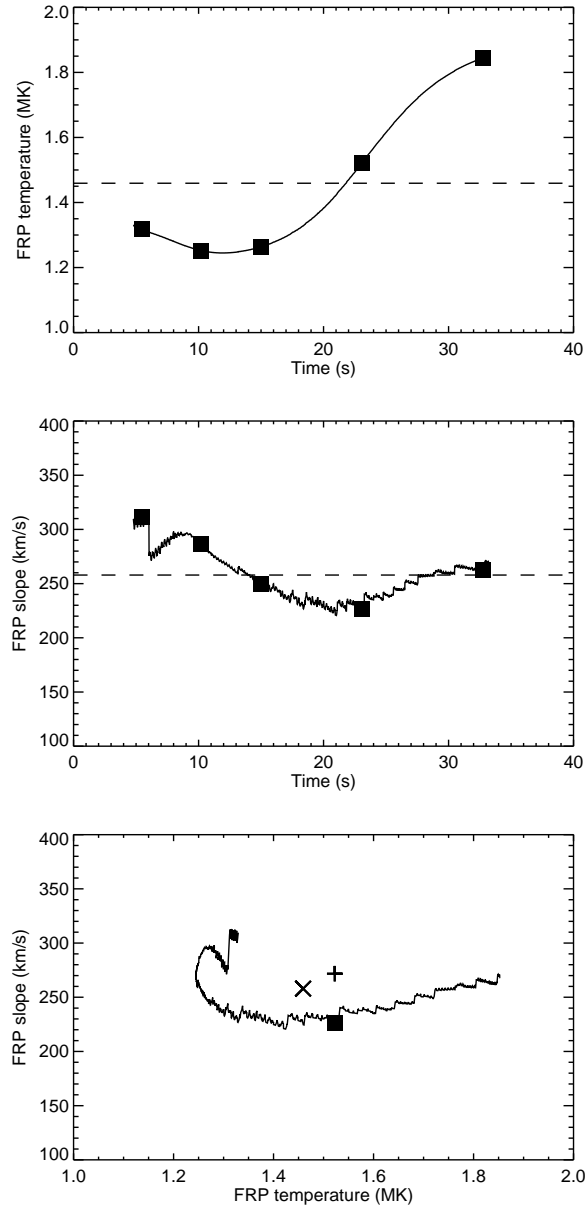}
\caption{Upper and middle plots: Time evolution of the FRP temperature and slope, respectively, as determined from the synthetic Doppler velocities in Section \ref{sec:wtdveltemp}.  The solid squares indicate the times shown in Figure \ref{fig:veltemp}, and the dashed lines indicate the mean value for each.  Lower plot: Evolution of the FRP slope plotted versus temperature.  The ``+'', solid square, and ``$\times$'' respectively indicate the observed, synthetic, and mean synthetic FRP properties from Sections \ref{sec:wtdveltemp} \& \ref{sec:fit}.
\label{fig:fcp_evolution}}
\end{figure}

In the lower plot of Figure \ref{fig:veltemp} we have indicated $\widetilde{T}_{frp}$ and $\widetilde{S}_{frp}$ for the $t=32.7$ sec profile as the vertical dashed line and the dash-dotted line respectively.  As an inspection of this plot will indicate, however, these two quantities do vary over the course of the simulation.  To track the evolution of $\widetilde{T}_{frp}$ and $\widetilde{S}_{frp}$ we have plotted them as functions of time in the upper and middle plots in Figure \ref{fig:fcp_evolution}, respectively, and as functions of each other in the lower plot in Figure \ref{fig:fcp_evolution}.  The notable ``jitteriness'' of the $\widetilde{S}_{frp}$ profile is due to movement between temperature bins when tracking $\widetilde{v} \ge 0.01$.  We have also plotted the values for $\widetilde{T}_{frp}$ and $\widetilde{S}_{frp}$ at the five times shown in Figure \ref{fig:veltemp} as solid squares in the upper two plots.  With some exceptions (discussed in Section \ref{sec:powerlaws}) the behavior of the FRP properties for other simulations is similar to that seen in Figure \ref{fig:fcp_evolution}.

Finally, we define a mean FRP temperature $\langle \widetilde{T}_{frp} \rangle$ and slope $\langle \widetilde{S}_{frp} \rangle$, calculated by taking a time-average of $\widetilde{T}_{frp}(t)$ and $\widetilde{S}_{frp}(t)$ over the range $(W/2) \le t \le (t_{sim} - W/2)$.  For this simulation $\langle \widetilde{T}_{frp} \rangle = 1.46$ MK and $\langle \widetilde{S}_{frp} \rangle = 260$ km s$^{-1}$, and we have plotted these values in Figure \ref{fig:fcp_evolution} as the horizontal dashed lines in the top and middle plots for $\widetilde{T}_{frp}$ and $\widetilde{S}_{frp}$, respectively, and as an $\times$ in the lower plot.  Obviously, the time-dependent values for the FRP properties differ from these mean values over the course of the simulation; in fact, the lower plot in Figure \ref{fig:fcp_evolution} shows that they do not ever assume the mean values simultaneously.  However, by using the mean values we can in some sense characterize the entire evolution of $\widetilde{T}_{frp}$ and $\widetilde{S}_{frp}$ for a given simulation, which will allow us in Section \ref{sec:powerlaws} to directly compare these values for many different simulations.

\subsection{Observational Data Fit}
\label{sec:fit}

\begin{figure}
\epsscale{0.5}
\plotone{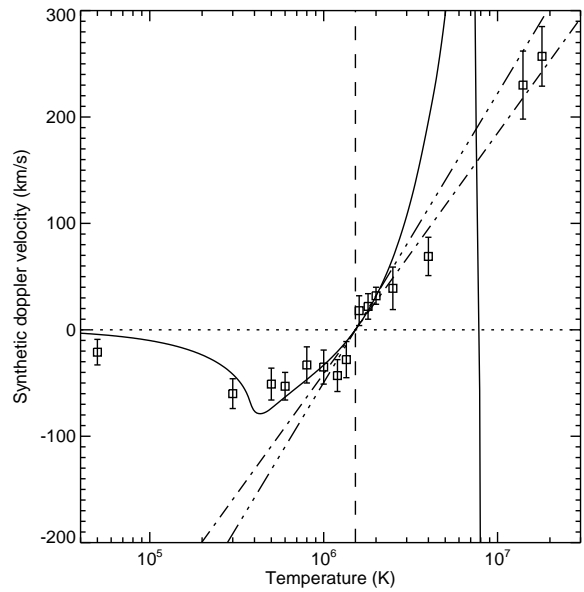}
\caption{Best fit of the synthetic Doppler velocity from the simulation considered in Section \ref{sec:sim} to the observed chromospheric Doppler velocities presented in \cite{milligan11}, as described in Section \ref{sec:fit}.  Squares and error bars are the \cite{milligan11} data (with reversed signs), and the dashed line and dash-triple-dotted lines are the approximate FRP temperature and slope, respectively.  The solid line is the best fit of the synthetic Doppler velocity, while the dash-dotted line is the synthetic FRP slope.
\label{fig:milligan_simfit}}
\end{figure}

We conclude our discussion of this particular simulation by making a comparison of our synthetic Doppler velocities to the observed flare loop Doppler velocities published by \cite{milligan09} (with a correction published in \cite{milligan11}).  The event studied in that paper was a \emph{GOES}-class C1.1 flare that took place in NOAA AR 10978 on 2007 December 14 at 14:12 UT.  Serendipitously, the \emph{Hinode/EIS} instrument was rastering over one of the flare loop footpoints during the impulsive phase of the flare, with an exposure time of 10 seconds.  The authors used 15 different spectral lines, with formation temperatures ranging from $5 \times 10^4$ K to 16 MK, to derive Doppler velocities for the footpoint plasma.  We have taken these Doppler velocity data and associated error ranges from Table 1 in \cite{milligan11}, and replotted them in Figure \ref{fig:milligan_simfit} as the square points and error bars (note that we have changed the signs on these data to match our velocity convention).

Inspection of the observational data in Figure \ref{fig:milligan_simfit} reveals similar structure to the synthetic Doppler velocity profiles in Figure \ref{fig:veltemp}, with downflows and upflows separated by a flow reversal temperature that lies somewhere between 1-2 MK.  To estimate the FRP parameters for comparison to the simulation, we take the six velocity measurements that fall in the 1-2 MK range, and use a linear fit of the form
\begin{equation}
v \left( T \right) = \widehat{C}_0 + \widehat{S}_{frp} \log_{10} T,
\end{equation}
where $\widehat{S}_{frp}$ is the approximate FRP slope; the approximate FRP temperature is given by $\widehat{T}_{frp} = 10^{-\widehat{C}_{0}/\widehat{S}_{frp}}$.  For the \cite{milligan11} data we find that $\widehat{T}_{frp} = 1.5$ MK and $\widehat{S}_{frp} = 270$ km s$^{-1}$, and these values are represented by the ``$+$'' in Figure \ref{fig:fcp_evolution} as well as the vertical dashed line and dash-triple-dotted line, respectively, in Figure \ref{fig:milligan_simfit}.

Since the observed flow conversion temperature is the quantity in which we have the most confidence, we fit the simulation to the data by selecting the time for which $\widetilde{T}_{frp}$ provides the best match to $\widehat{T}_{frp}$.  This is found to be at $t=23.1$ sec (orange line in Figures \ref{fig:sim_evolution}-\ref{fig:veltemp}) with $\widetilde{T}_{frp} = 1.5$ MK and $\widetilde{S}_{frp} = 230$ km s$^{-1}$, plotted as the solid square in Figure \ref{fig:fcp_evolution}.  That profile for the synthetic Doppler velocity has been plotted on top of the observation data in Figure \ref{fig:milligan_simfit}, along with a dash-dotted line representing the slope $\widetilde{S}_{frp}$.  Aside from the approximate match to the FRP properties, we note that the profile from the simulation matches well to the observed velocities in the range $5 \times 10^5$ K $\le T \le 2 \times 10^6$ K.  Outside of this range, however, the simulation profile begins to diverge from the observed values, especially for the highest temperatures (12.5 MK and 16 MK).  Indeed, these temperatures do not even exist in this simulation, which has a maximum post-shock temperature of 9.2 MK.

We also observe from Figure \ref{fig:fcp_evolution} that the mean slope $\langle \widetilde{S}_{frp} \rangle$ is a somewhat better estimate for $\widehat{S}_{frp}$ than the value for $\widetilde{S}_{frp}$ at $t=23.1$ secs, but that $\langle \widetilde{T}_{frp} \rangle$ is a slight underestimate for $\widehat{T}_{frp}$.  Nevertheless, as an order-of-magnitude estimation, the mean value is not far removed from the observational result.  Thus, we feel we are justified in using the mean FRP properties $\langle \widetilde{T}_{frp} \rangle$ and $\langle \widetilde{S}_{frp} \rangle$ as proxies for describing the overall evolution of the evaporation during a flare.

\section{Scaling Laws}
\label{sec:powerlaws}

Thus far we have developed a method for reducing the complex properties of the spatially- and temporally-dependent chromospheric evaporation driven by thermal conduction in a simple model atmosphere down to two scalar quantities, namely $\langle \widetilde{T}_{fc} \rangle$ and $\langle \widetilde{S}_{fc} \rangle$.  Further, we have shown that these quantities provide an acceptable description of the FRP properties in observed chromospheric flows.  An obvious question now presents itself: is it possible to systematically relate the observed FRP properties $\widehat{T}_{fc}$ and $\widehat{S}_{fc}$ back to fundamental parameters in the simulation?

Of course, with only two data inputs, it will only be possible to extract at most two parameters for the simulation.  The two most useful properties are likely the TR temperature ratio $R$, which gives some insight into the pre-flare state of the loop, and the Mach number of the piston shock $M_p$, which has been shown previously to relate to the initial reconnection angle of the loop \citep{longcope09}.  We therefore seek an invertible relationship between $( \langle \widetilde{T}_{frp} \rangle, \langle \widetilde{S}_{frp} \rangle )$ and $( M_p, R )$; as we shall show, such a relationship does exist, but instead of $M_p$ we shall use the post-shock temperature $T_{ps}$.  However, the post-shock temperature is dictated uniquely by the piston Mach number, as described in Section \ref{sec:shock}, and thus the two are effectively equivalent.

To determine the desired parameter relationships, we require a set of simulations that adequately span the parameter space.  At this point, in order to avoid confusion, we shall begin labeling dimensionless quantities explicitly with a superscript ``$*$''; variables with ordinary units will be left as normal.  We choose five values for the piston Mach number $M_p^*$, given by
\begin{equation}
M_p^* = \left( 2.0, \, 2.5, \, 3.0, \, 3.5, \, 4.0 \right);
\end{equation}
these values translate to the equivalent (dimensionless) post-shock temperatures,
\begin{equation}
T_{ps}^* = \left( 3.67, \, 4.93, \, 6.47, \, 8.28, \, 10.4 \right).
\end{equation}
We also choose five values for the TR temperature ratio $R^*$, given by
\begin{equation}
R^* = \left( 100, \, 150, \, 200, \, 250, \, 300 \right).
\end{equation}
Simulations were performed for all 25 combinations of the two parameters.  For brevity, we will label and discuss these simulations based on the parameter values selected: the five values for $T_{ps}^*$ are labeled ``A''-``E'', and the five values for $R^*$ are labeled ``1''-``5''.  Thus, the simulation discussed at length in Section \ref{sec:sim}, which had $M_p=2.0$ and $R=250$ would be labeled as ``A4''.

\begin{figure}
\epsscale{1.0}
\plotone{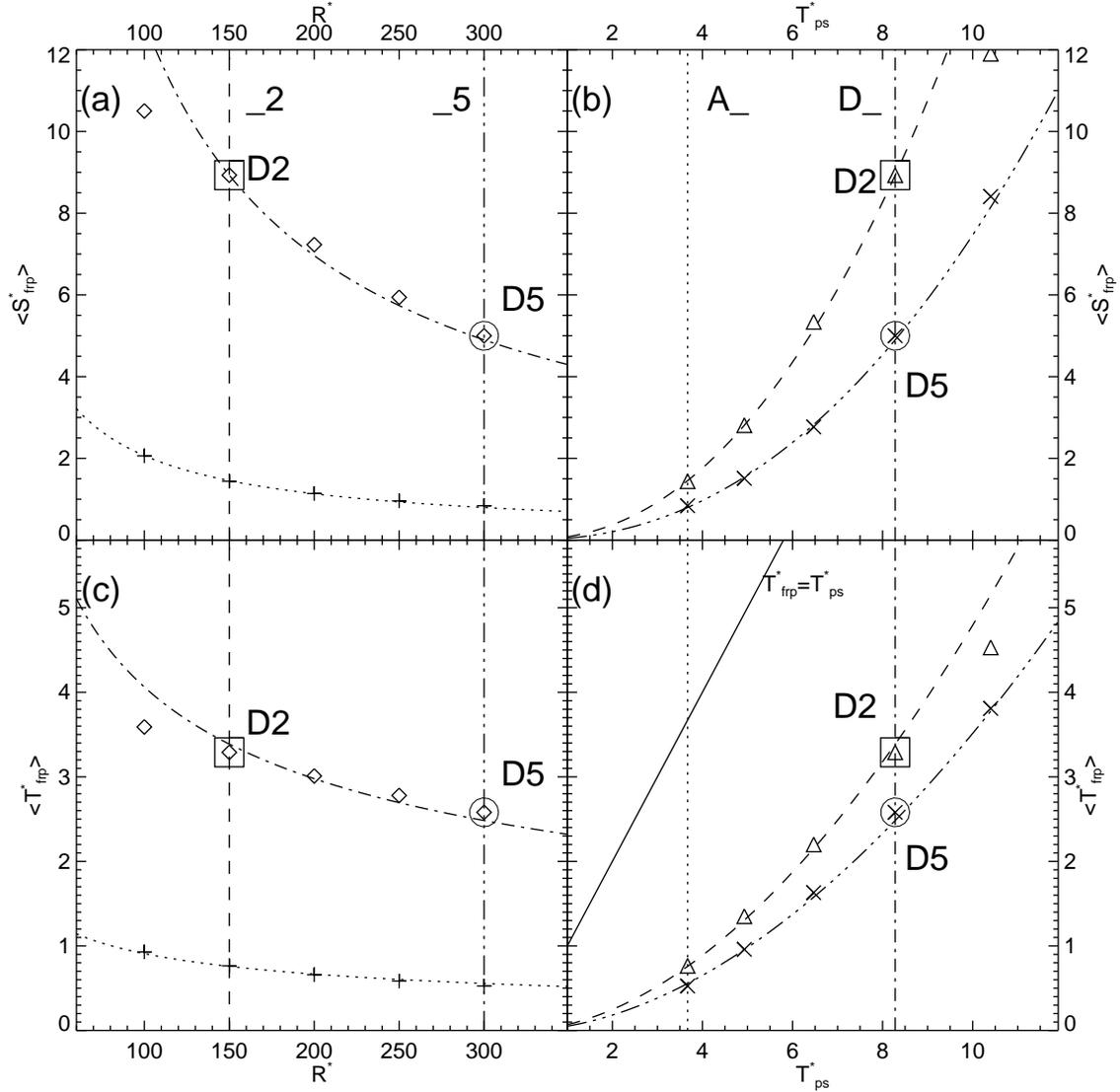}
\caption{Some of the runs conducted for the parameter survey.  Four of the 10 sequences are shown differentiated by line style: ``2'' (dashed), ``5'' (broken), ``A'' (dotted) and ``D'' (dash-dot).  These lines show the power-law fits from Equations (37) and (38).  The data points themselves are shown by symbols.  The left and right columns show the variation in run conditions $R^*$ and $T_{ps}^*$.  The top and bottom rows show the variation in measured characteristics, $\widetilde{S}_{frp}^*$ and $\widetilde{T}_{frp}^*$.  To indicate the relation between the panels runs D2 and D5 are indicated by text and by larger squares and circles respectively.
\label{fig:survey}}
\end{figure}

We next calculate, as in Section \ref{sec:wtdveltemp}, the dimensionless mean synthetic Doppler velocity FRP temperature $\langle \widetilde{T}^*_{frp} \rangle = \langle \widetilde{T}_{frp} \rangle / T_{cor}$ and slope $\langle \widetilde{S}^*_{frp} \rangle = \langle \widetilde{S}_{frp} \rangle / c_{s,cor}$ for each simulation.  The values of these two quantities for all 25 simulations are tabulated in Table 1.  In Figure \ref{fig:survey} we show the results of sequences ``2'' (dashed), ``5'' (broken), ``A'' (dotted) and ``D'' (dash-dot).  This parameter survey shows that increasing the chromospheric density ratio $R^*$ leads to decreases in both FRP temperature (9c) and slope (9a).  Increasing the post-shock temperature $T_{ps}^*$ leads to an increase in each of these (9d and 9b).  The flow conversion temperature lies naturally beneath the post-shock temperature solid curve in Figure 9d but can be below the coronal temperature if the chromospheric ratio is sufficiently large ($T_{frp}^*=1$ in Figures 9c and 9d).  All 25 runs are fit to a pair of power law relations of the form
\begin{equation}
\langle \widetilde{T}^*_{frp} \rangle = C_1 \left(T_{ps}^*\right)^{A_{11}} \left(R^*\right)^{A_{12}},
\label{eq:nondimT}
\end{equation}
\begin{equation}
\langle \widetilde{S}^*_{frp} \rangle = C_2 \left(T_{ps}^*\right)^{A_{21}} \left(R^*\right)^{A_{22}},
\label{eq:nondimS}
\end{equation}
where $A_{11} = 1.84$, $A_{12} = -0.448$, $A_{21} = 2.23$, $A_{22} = -0.866$, $C_1 = 0.654$, and $C_2 = 6.14$.  These fits are plotted as curves in Figure \ref{fig:survey}.

For the foregoing power-law fits we have chosen to omit the values for the ``D1'', ``E1'', and ``E2'' simulations, as doing so results in a much stronger fit for the remaining 22 simulations (the percent error between the actual values and the fit values are also tabulated in Table 1).  As shown in Figure \ref{fig:survey}, the three omitted simulations have values for $\langle \widetilde{T}^*_{frp} \rangle$ that fall between 10-19\% below the power-law fit, and values for $\langle \widetilde{S}^*_{frp} \rangle$ that fall between 14-38\% below the power-law fit.  We believe the reason for the poor fit for these simulations is that they fall in a ``weak-TR/strong-shock'' regime, where the TCF simply moves through the TR and chromosphere too quickly, and thus the time-averaged FRP properties do not reflect the later evolution seen in Figure \ref{fig:fcp_evolution}.

Since the dimensionless versions of the scaling laws (\ref{eq:nondimT}) \& (\ref{eq:nondimS}) are not especially useful for handling observational data, we rescale them by using the definitions $R^* = T_{cor}/T_{chr}$ and $T^*_{ps} = T_{ps}/T_{cor}$, and the definitions of $\langle \widetilde{T}^*_{frp} \rangle$ and $\langle \widetilde{S}^*_{frp} \rangle$ given above.  We also use the fact that the sound speed in the corona is given by $c_{s,cor} = c_{s,0} T_{cor}^{1/2}$, where $c_{s,0} = 0.17$ km s$^{-1}$ K$^{-1/2}$.  After some rearrangement, and making the assumption that $\widehat{T}_{frp}$ and $\widehat{S}_{frp}$ are adequate proxies for $\langle \widetilde{T}^*_{frp} \rangle$ and $\langle \widetilde{S}^*_{frp} \rangle$, we obtain the following:
\begin{equation}
\widehat{T}_{frp} = C_1 \, T_{chr}^{-A_{12}} \, T_{ps}^{A_{11}} \, T_{cor}^{A_{12}-A_{11}+1}
\label{eq:scalingT}
\end{equation}
\begin{equation}
\widehat{S}_{frp} = C_2 \, c_{s,0} \, T_{chr}^{-A_{22}} \, T_{ps}^{A_{21}} \, T_{cor}^{A_{22}-A_{21}+1/2}.
\label{eq:scalingS}
\end{equation}
Note that these versions of the scaling relationships require the assumption of one of the parameters $T_{chr}$, $T_{cor}$, or $T_{ps}$.  Recall that throughout Section \ref{sec:sim} we restored the simulation variables to conventional units in part by setting $T_{cor} = 2.5 \times 10^{6}$ K.  Since $R^* = 250$ for that simulation, it follows that $T_{chr} = 10^4$ K.  We now generalize this assumption for all our simulations by fixing $T_{chr} = 10^4$ K, and we then invert Equations (\ref{eq:scalingT}) \& (\ref{eq:scalingS}) to yield the flare parameters explicitly from observables:
\begin{eqnarray}
  T_{ps} &=& G_1\, T_{chr}^{D_{1}}\, \widehat{T}_{frp}^{B_{11}}\,
  \widehat{S}_{frp}^{B_{12}} \label{eq:invertT} \\
  T_{cor} &=& G_2\, T_{chr}^{D_{2}}\, \widehat{T}_{frp}^{B_{21}}\,
 \widehat{S}_{frp}^{B_{22}} \label{eq:invertS}
\end{eqnarray}
where $B_{11}=1.36$, $B_{12}=-0.678$, $B_{21}=1.17$, $B_{22}=-0.967$, $G_1=1.84$, $G_2=1.71$, $D_{1} = B_{11}A_{12}+B_{12}A_{22} = -0.0225$ and $D_{2} = B_{21}A_{12}+B_{22}A_{22} = 0.313$, and $\widehat{S}_{frp}$ is in units of km s$^{-1}$.

As a final check we determine if the run presented in Section \ref{sec:sim}, which fit both the observed Doppler velocity data and flow conversion properties quite well, is actually the run that would be suggested by the above scaling laws.  We adopt $T_{chr} = 10^4$ K and use the observed values for $\widehat{T}_{frp} = 1.5$ MK and $\widehat{S}_{frp} = 270$ km s$^{-1}$ in Equations (\ref{eq:invertT}) \& (\ref{eq:invertS}) to obtain a suggested coronal temperature $T_{cor} = 2.4$ MK and post-shock temperature $T_{ps} = 8.4$ MK, implying a dimensionless TR ratio $R^* = 240$ and post-shock temperature $T^*_{ps} = 3.5$.  From Table 1, we see that the simulation closest to these suggested values is indeed ``A4'', with $R^* = 250$ and $T^*_{ps} = 3.67$, indicating that the derived scaling laws do indeed yield reasonable results.

\section{Discussion}
\label{sec:discussion}

In this paper we have developed a numerical simulation code to investigate relationships between certain observable properties of chromospheric evaporation and more fundamental (but difficult to determine) properties of the loop atmosphere and of coronal energy release.  This code is an extension of the model developed in \cite{longcope09}, in which the coronal plasma in a post-reconnection flux tube is accelerated and compressed due to the contraction of the loop under magnetic tension.  The compressed plasma at the loop top results in slow magnetosonic shocks, and the post-shock plasma is heated to flare temperatures of several megakelvin from the conversion of free magnetic energy to kinetic and then to thermal energy.  We have extended this model to include a highly simplified model chromosphere and transition region (TR) at the loop footpoints to act as a mass reservoir.  Thermal conduction from the post-shock plasma transports heat down toward the footpoints, resulting in impulsive heating of the chromospheric and TR plasma and a disruption of the local thermodynamic equilibrium and finally in the bulk flows of plasma known as chromospheric evaporation and condensation.

In creating the numerical simulation code, we made a variety of simplifying assumptions beyond the piston shock model.  First, we opted to ignore gravitational stratification and loop geometry, effectively assuming a horizontal tube of uniform cross-section.  We also left out explicit radiation and coronal heating, instead setting up our model loop atmosphere using a simple function for the temperature profile.  We then calculated the necessary heating input to maintain that atmosphere at equilibrium and supplied that to the loop for the entire simulation.  We chose these particular simplifications in part because they allowed us to assign definite values for certain parameters of interest (e.g.\ the ratio of the coronal to chromospheric temperatures), and thus more easily make comparisons between simulations with different values for those parameters.  We did test the effect of the heating input on the simulations by turning it off for a test run, and we found no significant impact on the result for that simulation.  Although we did not present it, we also conducted a preliminary test varying the thickness of the transition region, and found only a very weak impact on the results.  However, since we used the same function for the temperature profile for all simulations, we are unable to make any claims about how a different temperature and density profile might alter the results.  Finally, recall that in order to adequately resolve features in the chromosphere, we artificially enhanced the viscosity and conductively at low temperatures.  This has the effect of altering short-scale physics, and also reducing velocities at low-temperatures, and decreasing this enhancement by 50\% results in 20\% and 10\% increases in $T_{frp}$ and $S_{frp}$, respectively.  We do not believe, however, that this substantially alters either our data comparison or our final conclusions.

To demonstrate the results of the code, we discussed in detail the hydrodynamic and DEM evolution of a single simulation, as the results for our other simulations were qualitatively similar and differed only in quantitative details.  In terms of broad-scale features, we observe that our results using this code are also qualitatively similar to simulations performed in other studies of impulsive conduction-driven evaporation.  This includes the rapid development of the thermal conduction front (TCF) and development of the isothermal subshock (ISS), the overpressure in the TR, and the presence of both upflows and downflows (so-called ``explosive'' evaporation) with upflow velocities dominating \citep{nagai80,cheng84,macneice86,fisher86}.  The magnitude and shape of the DEM profile was found to be similar to other studies \citep{emslie85,brosius96}.  This is encouraging, since our model is in many ways simpler than other models (as discussed above) but still seems to evolve in a broadly similar manner.  Our model differs from most others, however, in one particularly substantial aspect: the use of a hydrodynamic shock to drive the TCF.  This shock, and the associated downflows, introduce a complex interplay between the evaporation front and the isothermal subshock.  It also serves to resolve the issue of the free-streaming saturation limit, which is easily violated using \emph{ad hoc} heating models.  Instead we find that, after a brief initial violation in the piston shock, the decomposition into the TCF and ISS quickly restores the thermal flux to below the saturation limit.

Our goal in this study was not to investigate the hydrodynamic and DEM evolution of chromospheric evaporation, which, as we have noted above, have been extensively studied elsewhere and in greater depth.  Instead, our purpose was to draw a connection to footpoint velocities obtained from spectral Doppler shifts in data from instruments such as \emph{Hinode/EIS}.  To do this, we constructed ``synthetic'' Doppler velocities from our simulation results by first weighting the plasma flow velocities by the emission measure, then binning over temperature, and finally averaging over a time-window of similar duration to typical \emph{Hinode/EIS} exposure times.  Although this method is not as exact as, for example, using the CHIANTI database to construct the actual spectral lines and associated Doppler shifts, we find that the results are more flexible for investigating temperature-dependent properties of observable footpoints flows.

The properties of particular note for our purpose center on the existence of a \emph{flow reversal point} (FRP) near the loop footpoint that separates upflows from downflows.  As seen in observational data, the FRP occurs at an identifiable temperature and with an identifiable velocity-temperature slope, and these properties can be seen to vary over time and among flares \citep{milligan09,milligan11,li11,raftery09}.  We have identified and tracked these two properties for our synthetic Doppler data, and we find that they do indeed vary over time, but that they are also confined to a fairly narrow range of values for each simulation.  We thus calculated the mean values for the FRP properties during each simulation to serve as a convenient proxy while comparing different simulations.  Finally, we compared the velocity data from \cite{milligan11} to both the ``best-fit'' and the mean FRP for one of the simulations and determined that both may be considered reasonable given the data.  This fact, along with the observation that the FRP properties evolve similarly across a range of simulation inputs, indicates that the mean FRP properties are a robust proxy for the overall FRP evolution during impulsive evaporation.

Having chosen the simulation inputs (namely the Mach number of the piston $M_p$ and the ratio of coronal-to-chromospheric temperatures $R$) and outputs (the mean FRP temperature and slope), we then investigated whether a simple relationship may be determined using this simulation code.  We have shown that it is possible to extract a general scaling-law relationship between these quantities, after replacing the piston shock Mach number with the (formally identical) post-shock temperature $T_{ps}$.  A few exceptions to these relationships exist, particularly among the simulations where the ratio $R$ is low and the Mach number $M_p$ is large (the ``strong-shock/weak-TR'' limit), but we note that the relationships hold well if these cases are ignored.  It is possible that these cases, which we note have the shortest duration among the 25 simulations, simply do not have enough time to evolve as fully as the others, particularly the FRP temperature which tends to rise as the simulations evolve.  One potential solution would be to extend the length of the simulation region, particularly the chromospheric depth, to allow for runs that are of equal duration; however, this has not been tested.  Finally, we observe that the scaling exponent trends make some physical sense: a larger value for $T_{ps}$ (i.e.\ a stronger driving shock) results in higher temperatures and flow velocities and hence in larger values for the mean FRP properties, whereas a larger value for $R$ (i.e.\ a hotter pre-flare corona) results in lower temperatures and slower flows and hence suppresses the mean FRP properties.  However at this point we have no intuition regarding the exact values for the scaling exponents, and further work will be needed to determine if these values may be derived analytically.

When applied to the data from \cite{milligan11}, the scaling-law relationships we have determined suggest input parameters that a reasonably close to the ``best-fit'' that we selected from among our performed simulations.  This again strongly suggests that the mean FRP properties are an adequate proxy for the overall evolution of the evaporation dynamics.  Further, the ambient coronal temperature and post-shock temperature are both reasonable and typical for active regions and flare loops.  That being said, we have also performed preliminary tests of our scaling law relationships on other observed flare footpoint flow profiles (e.g.\ \cite{li11}, \cite{raftery09}) and we find these results somewhat less encouraging.  For example, one set of Doppler velocities from \cite{li11} yield an ambient coronal temperature of $\sim$24 MK and post-shock temperature of $\sim$45 MK, and data from \cite{raftery09} suggests $T_{cor} \approx 8$ MK and $T_{ps} \approx 15$ MK.  In both cases the ambient coronal temperature is significantly higher than is typically expected, and the post-shock temperature suggests a very weak effective piston shock ($M_p < 2$).  Data from both papers suggest FRP temperatures comparable to what we have presented here, but have FRP slopes that are much shallower than is seen in any of our simulations.

It therefore seems likely that there is some effect, besides viscosity, acting to suppress the plasma velocities in at least the evaporation region, and possibly the condensation region as well.  We posit, without proof, that these unreasonable values for the coronal and post-shock temperature may be resolved by incorporating a flux tube ``nozzle'' at or near the TR.  Some constriction of the magnetic flux tube is expected near the loop footpoint due to a combination of canopy expansion and magnetic pressure.  The authors believe that the presence of a nozzle near the TR will modify the properties of the resulting chromospheric flows, possibly suppressing flow velocities in such a way that the shallow FRP slope is reproduced for more reasonable parameters of the flare loop.  This topic will be the subject of a future study using an extension of this numerical simulation model.

\acknowledgments

We wish the thank the referee for a careful reading of our manuscript and for providing constructive comments which improved the manuscript.  We also thank Dr.\ Sylvina Guidoni, Dr.\ Lucas Tarr, and Roger Scott for many productive discussions during the development of the code used in this paper, and also Prof.\ David McKenzie for his valuable input in preparing the manuscript.  This work was supported by a grant from the National Science Foundation (NSF) and Department of Energy (DOE) Partnership for Plasma Sciences and a NASA SR\&T grant.

\clearpage



\clearpage

\clearpage

\clearpage

\clearpage

\clearpage

\clearpage

\clearpage

\clearpage

\clearpage

\begin{table}
\begin{center}
\caption{Simulation labels and properties.\label{tbl-1}}
\begin{tabular}{crrrrrrrrrrr}
\tableline\tableline
Label & $R^*$ & $M^*_{p}$ & $T^*_{ps}$ & $\langle \widetilde{T}^*_{frp} \rangle$ & Fit error & $\langle \widetilde{S}^*_{frp} \rangle$ & Fit error & Fit? \\
\tableline
A1 & 100 & 2.0 & 3.67 & 0.929  & 1.8 \% & 2.06 & -1.0 \% & Yes \\
A2 & 150 & 2.0 & 3.67 & 0.764  & 0.36 \% & 1.44 & -1.8 \% & Yes \\
A3 & 200 & 2.0 & 3.67 & 0.659  & -1.5 \% & 1.14 & -0.38 \% & Yes \\
A4 & 250 & 2.0 & 3.67 & 0.584  & -3.5 \% & 0.960 & 2.1 \% & Yes \\
A5 & 300 & 2.0 & 3.67 & 0.526  & -5.7 \% & 0.841 & 4.8 \% & Yes \\
B1 & 100 & 2.5 & 4.93 & 1.60  & 1.8 \% & 4.13 & 2.7 \% & Yes \\
B2 & 150 & 2.5 & 4.93 & 1.35  & 2.7 \% & 2.81 & -0.63 \% & Yes \\
B3 & 200 & 2.5 & 4.93 & 1.18  & 2.1 \% & 2.15 & -2.5 \% & Yes \\
B4 & 250 & 2.5 & 4.93 & 1.05  & 1.1 \% & 1.76 & -2.9 \% & Yes \\
B5 & 300 & 2.5 & 4.93 & 0.959  & -0.19 \% & 1.51 & -2.9 \% & Yes \\
C1 & 100 & 3.0 & 6.47 & 2.51  & -3.2 \% & 7.42 & 0.49 \% & Yes \\
C2 & 150 & 3.0 & 6.47 & 2.20  & 1.5 \% & 5.34 & 2.8 \% & Yes \\
C3 & 200 & 3.0 & 6.47 & 1.96  & 2.9 \% & 4.08 & 0.74 \% & Yes \\
C4 & 250 & 3.0 & 6.47 & 1.77  & 3.0 \% & 3.29 & -1.5 \% & Yes \\
C5 & 300 & 3.0 & 6.47 & 1.63  & 2.7 \% & 2.77 & -2.8 \% & Yes \\
D1 & 100 & 3.5 & 8.28 & 3.59  & -12 \% & 10.5 & -18 \% & No \\
D2 & 150 & 3.5 & 8.28 & 3.29  & -3.7 \% & 8.93 & -0.96 \% & Yes \\
D3 & 200 & 3.5 & 8.28 & 3.01  & 0.38 \% & 7.23 & 3.0 \% & Yes \\
D4 & 250 & 3.5 & 8.28 & 2.78  & 2.4 \% & 5.94 & 2.6 \% & Yes \\
D5 & 300 & 3.5 & 8.28 & 2.58  & 3.3 \% & 5.00 & 1.1 \% & Yes \\
E1 & 100 & 4.0 & 10.4 & 4.78  & -23 \% & 11.3 & -47 \% & No \\
E2 & 150 & 4.0 & 10.4 & 4.53  & -13 \% & 11.9 & -21 \% & No \\
E3 & 200 & 4.0 & 10.4 & 4.27  & -6.5 \% & 11.0 & -6.2 \% & Yes \\
E4 & 250 & 4.0 & 10.4 & 4.03  & -2.4 \% & 9.63 & -0.07 \% & Yes \\
E5 & 300 & 4.0 & 10.4 & 3.81  & 0.13 \% & 8.41 & 2.2 \% & Yes \\
\tableline
\end{tabular}
\end{center}
\end{table}
\clearpage


\end{document}